\documentclass[published]{JHEP3}
\JHEP{00(2009)000}

\usepackage{epsfig,multicol,bbm}

\usepackage{amsmath}

\def\beq{\begin{equation}}
\def\eeq{\end{equation}}
\def\bea{\begin{eqnarray}}
\def\eea{\end{eqnarray}}

\title{Unusual Thermodynamics on the Fuzzy 2-Sphere}

\author{ Sanatan Digal\\ The Institute of Mathematical Sciences, CIT
Campus, Taramani, Chennai 600 113, India \\ E-mail:
\email{digal@imsc.res.in}}

\author{Pramod Padmanabhan \\ Department of Physics, Syracuse University, Syracuse, NY
13244-1130, USA \\ The Institute of Mathematical Sciences, CIT
Campus, Taramani, Chennai 600 113, India  \\ E-mail:
\email{ppadmana@syr.edu}}

\received{\today} \accepted{\today}

\preprint{SU-4252-910}
\preprint{IMSc/2010/06/09}

\abstract
{Higher spin Dirac operators on both the continuum sphere($S^2$) and its
fuzzy analog($S^2_F$) come paired with anticommuting chirality
operators. A consequence of this is seen in the fermion-like
spectrum of these operators which is especially true even for the
case of integer-spin Dirac operators. Motivated by this feature of
the spectrum of a spin 1 Dirac operator on $S_F^2$, we assume the
spin 1 particles obey Fermi-Dirac statistics. This choice is inspite
of the lack of a well defined spin-statistics relation on a compact
surface such as $S^2$. The specific heats are computed in the cases
of the spin $\frac{1}{2}$ and spin 1 Dirac operators. Remarkably the
specific heat for a system of spin $\frac{1}{2}$ particles is more
than that of the spin 1 case, though the number of degrees of freedom
is more in the case of spin 1 particles. The reason for this is inferred through
a study of the spectrums of the Dirac operators in both the cases.
The zero modes of the spin 1 Dirac operator is studied as a
function of the cut-off angular momentum $L$ and is found to follow a simple
power law. This number is such that the number of states with positive
energy for the spin 1 and spin $\frac{1}{2}$ system become comparable.
Remarks are made about the spectrums of higher spin Dirac operators 
as well through a study of their zero-modes and the variation of their spectrum
with degeneracy. The mean energy as a function of temperature is studied in both
the spin $\frac{1}{2}$ and spin 1 cases. They are found to deviate from the
standard ideal gas law in 2+1 dimensions.}


\begin{document}
 
 \section{Introduction}
 The fuzzy 2-sphere($S_F^2$)~\cite{Madore, Fuzzybook} is an example
of a noncommutative spacetime where the algebra of functions on the
commutative 2-sphere($S^2$) is approximated by an associative matrix
algebra. In principle this helps in regularizing  field theories on
such spaces, providing an alternative way to discretize spacetimes.
The most non-trivial feature of this approach is the preservation of
symmetries of the commutative spacetime even at the discrete level.
Field theories on $S_F^2$ have been actively pursued in the last few
years~\cite{SbApbBySv, ApbSv, Hs, Ww2}. Several works on numerical
simulations of scalar fields and gauge fields on $S_F^2$ have also
been done.~\cite{Mp, Mp2, CrdSdTrg1,CrdSdTrg2, FgfXmDoc, DocBy}.
These works study phase transitions of scalar fields on the fuzzy
sphere and they show the existence of new phases in the continuum
which break translational as well as other global symmetries. The
study of a real scalar field with a $\phi^4$ interaction reveals the
existence of a non-uniform ordered phase which is similar to the
striped phase~\cite{Hs2, JaSc, WbFhJn}. This feature distinguishes
it from its commutative counterpart. Apart from these novel
phenomena one can also incorporate supersymmetry in a precise manner
on $S_F^2$~\cite{Fuzzysusy}.

Fuzzy spaces other than $S_F^2$ have also been seen in the context
of topology change~\cite{Beyondfuzzy, JaMbLhJhHs}. Physics on other
fuzzy spaces have also been considered~\cite{Cylinder,Torus}.

To add to these new features on $S_F^2$, in this paper we probe the
thermodynamics of spin 1 and spin $\frac{1}{2}$ particles on the
fuzzy sphere. We find several counterintuitive results which we will
present in this work. We work with the spectrum of the Dirac
operators for the spin 1 and spin $\frac{1}{2}$ case. This is
natural to do as the Dirac operator is fundamental to physics and is
useful in formulating metrical, differential geometric and
bundle-theoretic ideas. Moreover in Connes' approach to
noncommutative geometry~\cite{Connes}, the Dirac operator gains
fundamental significance as part of the spectral triple in
formulating the spectral action principle~\cite{AcAc}. There are
several ways to construct the Dirac operator on the fuzzy
sphere~\cite{Ww, HgCkPp, BdIhSmDc, Cj}. They construct the Dirac
operator for a spin $\frac{1}{2}$ particle. In our approach we
construct the Dirac operators using the Ginsparg-Wilson(GW)
algebra~\cite{PgKw}. Through this method we can extend the
construction to all higher spins as studied in~\cite{ApbPp}.

We consider the spin 1 Dirac operators constructed in~\cite{ApbPp}.
Unlike the spin $\frac{1}{2}$ case, analytic computation of the
spectrum of these operators are difficult. This lead us to compute
its spectrum numerically~\cite{SdPp}. Though we could not go to
arbitrarily large values of the cut-off angular momenta $L$,  we
could still predict the spectrum's behavior in the continuum by
observing the striking patterns that emerged for the spectrums for
the values we could compute for.  In~\cite{ApbPp}, 3 different Dirac
operators along with 3 chirality operators were constructed for the
spin 1 case. They are all unitarily inequivalent as proved
in~\cite{SdPp}. We consider only the spectrum of the traceless spin
1 Dirac operator in this work.

Using this spectrum we first compute the partition function for a
system of spin 1 particles on $S_F^2$. For doing this we need to
assume the particles obey a particular statistics. As we are dealing
with a chiral system we assume that the particles obey the
Fermi-Dirac statistics. However it should be noted that the
conventional proofs of the spin-statistics theorem hold  in
relativistic quantum field theories(qft's) in three or more
dimensions. They use the axioms of local relativistic qft's. For
comprehensive proofs see~\cite{Aws, DR}. Field theory on the fuzzy
sphere is not a relativistic one as the symmetry group of the
underlying theory is $SU(2)$. This being the case there is no well
defined spin-statistics relation on the fuzzy sphere. However there
are spin statistics relations which do not require relativity and
which are topological~\cite{ApbRsAmGm, Rt, DfJr}. General theory for
quantum statistics in 2 spatial dimensions have also been
discussed~\cite{Ysw}. The non-triviality in two spatial dimensions
arises due to topology of the configuration space of
indistinguishable particles living on such a space. The fundamental
group for such a configuration
space($\frac{(\mathbb{R}^2)^N-{\Delta}}{S_N}$, where
${\Delta}={(\vec{r}_1,\vec{r}_2,...,\vec{r}_N)|
\vec{r}_i\in\mathbb{R}^2 \textrm{and} ~\vec{r}_i=\vec{r}_j} $ for
some $i\neq j$ and $S_N$ is the symmetric group of $N$ particles.)
is the braid group $B_N$. For the case of $S^2$ instead of
$\mathbb{R}^2$, the fundamental group is still the braid group with
an additional constraint~\cite{Jb, ApbTrg}. These considerations
allow for the possibility of the assumption of anyonic
statistics~\cite{Ysw,Al,Fw,Ak} in our case, but we do not consider
these possibilities in this work and only briefly remark about them
in the final section.

The main result of this paper is the mean energy of the spin
$\frac{1}{2}$ system is more than that of the spin 1 system. This is
surprising given the fact that there are more number of states
possible in the spin 1 case, $3\times (2L+1)^2$ than in the spin
$\frac{1}{2}$ case, $2\times (2L+1)^2$. Nevertheless we come to
terms with this strange behavior by looking at the distribution of
the eigenvalues of the Dirac operators in both the cases. In
particular the spectrum of the spin 1 Dirac operator consists of a
number of zero modes which are absent in the spin
$\frac{1}{2}$ case. These characteristics provide an answer to the
strange behavior of the mean energies.  A consequence of this is
also seen in the specific heats of the two systems. The entropy of 
the spin $\frac{1}{2}$ system is also more than the spin 1 system.

The other result is the deviation of the plot of the mean energy vs
temperature from the corresponding curve for an ideal gas on a two
dimensional space. The mean energy of an ideal gas of massless
particles on a flat two dimensional space goes  as $T^3$. We find 
a deviation from this law which is attributed to the dispersion
relation of the spin $\frac{1}{2}$ and spin 1 systems on the 2-sphere.

The paper is organized as follows. Section 2 reviews the
noncommutative algebra of functions on $S_F^2$.  The GW algebra and
the construction of the Dirac operators are reviewed in section 3.
In section 4 we describe a way to find the spectrum of the spin 1 Dirac 
operator for arbitrarily large cut-off $L$. This is done by looking at
a particular scaling behavior found in their spectrum.
The grand canonical partition function is computed in section 5.
This is used to compute the mean energy and the specific heats for
both the cases of spin 1 and spin $\frac{1}{2}$ systems. This
section presents the relevant numerical results. Zero mode analysis
is also carried out in this section along with the reasons for the
strange behavior. We also speculate the behavior of the mean
energies for higher spin Dirac operators. Section 6 discusses the deviation from ideal gas
behavior of these systems. We conclude in
section 7 with a few remarks.

\section{Geometry of $S_F^2$}
The algebra for the fuzzy sphere is characterized by a cut-off
angular momentum $L$ and is the full matrix algebra $Mat(2L+1)\equiv
M_{2L+1}$ of $(2L+1)\times (2L+1)$ matrices. They can be generated
by the $(2L+1)$-dimensional irreducible representation (IRR) of
$SU(2)$ with the standard angular momentum basis. The latter is
represented by the angular momenta $L^L_i$ acting on the left on
$Mat(2L+1)$: If $\alpha\in Mat(2L+1)$,
\beq{L_{i}^{L}\alpha=L_{i}\alpha}\eeq
\beq{[L_{i}^{L},L_{j}^{L}]=i\epsilon_{ijk}L_{k}^{L}}\eeq
\beq{(L_{i}^{L})^2=L(L+1)\mathbf{1}}\eeq where $L_i$ are the
standard angular momentum matrices for angular momentum $L$.

We can also define right angular momenta $L_i^R$:
\beq{L_{i}^{R}\alpha=\alpha L_{i}, \alpha\in M_{2L+1}}\eeq
\beq\label{commu}{[L_{i}^{R},L_{j}^{R}]=-i\epsilon_{ijk}L_{k}^R}\eeq
\beq{(L_{i}^{R})^2=L(L+1)\mathbf{1}}\eeq We also have
\beq{[L_i^L,L_j^R]=0.}\eeq

The operator $\mathcal{L}_i=L_i^L-L_i^R$ is the fuzzy version of
orbital angular momentum. They satisfy the $SU(2)$ angular momentum
algebra
\beq{[\mathcal{L}_i,\mathcal{L}_j]=i\epsilon_{ijk}\mathcal{L}_k}\eeq

In the continuum, $S^2$ can be described by the unit vector
$\hat{x}\in S^2$, where $\hat{x}.\hat{x}=1$. Its analogue on $S_F^2$
is $\frac{L_i^L}{L}$ or $\frac{L_I^R}{L}$ such that
\beq{\lim_{L\rightarrow\infty}\frac{L_i^{L,R}}{L}=\hat{x}_i.}\eeq
This shows that $L_i^{L,R}$ do not have continuum limits. But
$\mathcal{L}_i=L_i^L-L_i^R$ does and becomes the orbital angular
momentum as $L\rightarrow\infty$:
\beq{\lim_{L\rightarrow\infty}L_i^L-L_i^R=-i(\overrightarrow{r}\wedge\overrightarrow{\nabla})_i
.}\eeq

\section{Construction of the Dirac Operators}
 In algebraic terms, the GW algebra $\mathcal{A}$ is the unital
$\ast$ algebra over $\mathbf{C}$ ,generated by two $\ast$-invariant
involutions $\Gamma, \Gamma'$.
 \beq \label{GW}\mathcal{A}=\{\Gamma,\Gamma'\ :\Gamma^2=\Gamma'^2=1\
,\Gamma^*=\Gamma\ ,\Gamma'^*=\Gamma'\}\eeq

 In any $\ast$ -representation on a Hilbert space,
$\ast$ becomes the adjoint $\dag$.

Consider the following two elements constructed out of $\Gamma,
\Gamma'$: \beq {\Gamma_1=\frac{1}{2}(\Gamma+\Gamma'),}\eeq
\beq{\Gamma_2=\frac{1}{2}(\Gamma-\Gamma').}\eeq It follows from
Eq.(\ref{GW}) that $\{\Gamma_1,\Gamma_2\}=0$. This suggests that for
suitable choices of $\Gamma$, $\Gamma '$, one of these operators may
serve as the Dirac operator and the other as the chirality operator
provided they have the right continuum limits after suitable
scaling.

 For the spin $1$ case the combination which leads to the desired
Dirac and chirality operators were found in~\cite{ApbPp} and they
are \beq\label{D1}{ D_1 =
L\left(\frac{\Gamma_{L+1}^L-\Gamma_{L-1}^R}{2}\right),}\eeq
\beq\label{D2}{ D_2 =
L\left(\frac{\Gamma_{L-1}^L-\Gamma_{L+1}^R}{2}\right)}\eeq and
\beq\label{D3}{ D_3 =
L\left(\frac{\Gamma_{L}^L-\Gamma_{L}^R}{2}\right).}\eeq with
\beq\label{C1}{\gamma_1 =
\left(\frac{\Gamma_{L+1}^L+\Gamma_{L-1}^R}{2}\right),}\eeq
\beq\label{C2}{\gamma_2 =
\left(\frac{\Gamma_{L-1}^L+\Gamma_{L+1}^R}{2}\right)}\eeq and
\beq\label{C3}{\gamma_3 =
\left(\frac{\Gamma_{L}^L+\Gamma_{L}^R}{2}\right)}\eeq as their
corresponding chirality operators. In the above equations
\beq\label{Gl+1l}{\Gamma_{L+1}^{L}=\frac{2(\vec{\Sigma}.\vec{L}^L+L+1)(\vec{\Sigma}.\vec{L}^L+1)-(L+1)(2L+1)}{(L+1)(2L+1)},}\eeq
\beq\label{Gl+1r}{\Gamma_{L+1}^{R}=\frac{2(-\vec{\Sigma}.\vec{L}^R+L+1)(-\vec{\Sigma}.\vec{L}^R+1)-(L+1)(2L+1)}{(L+1)(2L+1)},}\eeq
\beq\label{Gl-1l}{\Gamma_{L-1}^{L}=\frac{2(\vec{\Sigma}.\vec{L}^L-L)(\vec{\Sigma}.\vec{L}^L+1)-L(2L+1)}{L(2L+1)},}\eeq
\beq\label{Gl-1r}{\Gamma_{L-1}^{R}=\frac{2(\vec{\Sigma}.\vec{L}^R+L)(\vec{\Sigma}.\vec{L}^R-1)-L(2L+1)}{L(2L+1)},}\eeq
\beq\label{Gll}{\Gamma^L_L=\frac{-2(\vec{\Sigma}.\vec{L}^L-L)(\vec{\Sigma}.\vec{L}^L+L+1)-L(L+1)}{L(L+1)},}\eeq
and
\beq\label{Glr}{\Gamma^R_L=\frac{2(\vec{\Sigma}.\vec{L}^R+L)(-\vec{\Sigma}.\vec{L}^R+L+1)-L(L+1)}{L(L+1)}.}\eeq
The operators in Eq.(\ref{Gl+1l})-Eq.(\ref{Glr}) are generators of
GW algebras and are obtained from left and right projectors to
eigenspaces of the total angular momentum, $\vec{L}+\vec{\Sigma}$,
where $\vec{\Sigma}$ are the matrices representing the spin $1$
representation of $SU(2)$.

 The continuum limits of Eq.(\ref{D1})-Eq.(\ref{D3}) are
\beq\label{CD1}{D_{1}=(\vec{\Sigma}.\vec{\mathcal{L}}-(\vec{\Sigma}.\hat{x})^{2}+2)+2(\vec{\Sigma}.\hat{x})+\{\vec{\Sigma}.\vec{\mathcal{L}},\vec{\Sigma}.\hat{x}\},}\eeq
\beq\label{CD2}{D_{2}=(\vec{\Sigma}.\vec{\mathcal{L}}-(\vec{\Sigma}.\hat{x})^{2}+2)-2(\vec{\Sigma}.\hat{x})-\{\vec{\Sigma}.\vec{\mathcal{L}},\vec{\Sigma}.\hat{x}\}}\eeq
and
\beq\label{CD3}{D_{3}=\vec{\Sigma}.\vec{\mathcal{L}}-(\vec{\Sigma}.\hat{x})^{2}+2.}\eeq

 The corresponding chirality operators in the continuum are \beq\label{CC1}{\gamma_{1}=(\vec{\Sigma}.\hat{x})^{2}+(\vec{\Sigma}.\hat{x})-1,}\eeq
\beq\label{CC2}{\gamma_{2}=(\vec{\Sigma}.\hat{x})^{2}-(\vec{\Sigma}.\hat{x})-1}\eeq
and \beq\label{CC3}{\gamma_{3}=1-2(\vec{\Sigma}.\hat{x})^{2}}\eeq
respectively.

\section{Spectrum of the Spin 1 Dirac operator for large $N$}

For a given cut-off $L=(N-1)/2$, the eigenvalues of the spin $\frac{1}{2}$ Dirac operator
is given by $\pm(j+\frac{1}{2})$ where $j$ is the eigenvalue of the total
angular momentum. The degeneracy $(g)$ of each of these eigenvalues $(E)$ is given by 
\beq{g=2j+1.}\eeq This gives the relation between the energy eigenvalue and its degeneracy
as \beq\label{evghalf}{E_{\frac{1}{2}}=\pm \frac{g}{2}.}\eeq In contrast, the spin 1 system
is richer with varied eigenvalues for each $L$ including a number of
zero modes. Though we cant find the corresponding relation between the energy eigenvalue $E$ and
its degeneracy $g$ analytically we can do so numerically~\cite{SdPp}. 

The numerical results show that $g$ increases with $E$ for small values of $E$. After reaching a 
maximum for an intermediate value of $E$, $g$ decreases with $E$.
This is unlike the spin $\frac{1}{2}$ case where $g$ is linear in $E$.
After rotation (around origin, by a fixed angle) the plot of $E^\prime$ vs $g^\prime$ fits perfectly 
with a parabola. The rotation angle was found to be cut-off independent.
Encouraged by this we tried to find if there is any universality in $E^\prime$ vs $g^\prime$.
We found that results for different cut-off lie on a universal curve. This is done
by scaling $E'$ and $g'$ so that the minimum of different curves match. In figure.1 one can see
the results for $N=21,27$ and $45$ lie on top of each other after scaling. This result
strongly suggest that we can obtain the spectrum for arbitrarily larger cut-off $(N)$ if
we know the scaling variables for different $N$~\cite{SdPp}. The universality also suggests that
$E'$ vs $g'$ is described by only two independent parameters. Once we know
$E'$ vs $g'$ we rotate it to get the relation between $E$ and $g$, which is given by

\beq\label{evgone}{E_1= \frac{\sqrt{b(g)^2 - 4ac(g)}-b(g)}{2a}}\eeq where \beq{a=\alpha\cos ^2\theta,}\eeq 
\beq{b(g)=2\alpha\cos\theta(\sin\theta g+\eta) + \sin\theta,}\eeq 
\beq{c(g)=\alpha(\sin\theta g+\eta)^2+\beta - \cos\theta g}\eeq and \beq{ \theta = 2.26159~\textrm{radians}.}\eeq

As discussed above only $\beta (N)$ and $\eta (N)$ are independent parameters. For $\alpha$ we need to
know its value for a particular $N$. For all other values of $N$, $\alpha$ can be calculated using $\beta (N)$
and $\eta (N)$.

 \begin{figure}
\begin{center}
\mbox{
    \leavevmode
    { 
      \includegraphics[height=2.5in,width=4.5in, angle=0]{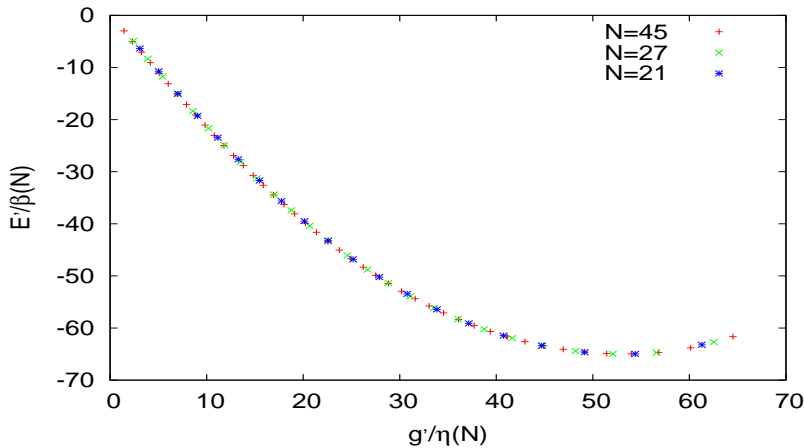} }}
 \end{center}
     \caption{Scaling property of the spectrum of the spin 1 Dirac operator.}
\end{figure}

\section{Numerical Results: I}

\subsection{The Partition Function and the Mean Energy}
As explained in the introduction, we assume the spin 1 particles to obey
fermionic statistics. The grand-canonical partition function is given by
\beq\label{part}{ln Z = \sum_{i=1} g_i ln\left(1+e^{-\beta (E_i-\mu)}\right)}\eeq
where $g_i$ is the degeneracy of the ith level, $E_i$ is the energy of the ith
level, $\mu$ is the chemical potential and $\beta=\frac{1}{k_BT}$. In the commutative case when there is no cut-off,
the product over $i$ extends till infinity, but here we are restricted by the
cut-off angular momentum $L$.

For the spin 1 case we numerically computed the spectrum of the
Dirac operator given by Eq.(\ref{D3}) in~\cite{SdPp}. We do not know how to find an 
analytic expression for the spectrum and so we compute the partition
function numerically. (See however~\cite{SdPp} for an analytic expression for the spectrum
of the spin 1 Dirac operator derived as a result of the numerical computations.) The analogous
situation for the spin $\frac{1}{2}$ case is fa better as we know its complete spectrum analytically
for arbitrarily large cut-off $L$. 

From the grand-canonical partition function in Eq.(\ref{part}) we
can use the standard formula to compute the mean energy which is
\beq\label{spin1en}{\langle E_1\rangle = -\frac{\partial ln
Z}{\partial\beta} = \sum_{i=1}^{(2L+1)^2-1} \frac{g_i
E_i}{e^{\beta(E_i-\mu)}+1}.}\eeq In what follows we take the
Boltzmann constant $k_B=1$ and the chemical potential $\mu=0$. The
above expression for the mean energy is used for both the spin 1 and
the spin $\frac{1}{2}$ cases. For the spin $\frac{1}{2}$ case it
becomes \beq\label{spin1/2en}{\langle E_{\frac{1}{2}}\rangle =
\sum_{j=1}^{2L-\frac{1}{2}} \frac{1}{2}\frac{(2j+1)^2}{e^{\beta(j+\frac{1}{2})}+1}.}\eeq Note
that in the above Eqns.(\ref{spin1en}, \ref{spin1/2en}) the sums are restricted by the cut-off $L$. 
The mean energies for both the cases were computed for various
temperatures from 0.1 to 50. We show only these plots here though we
did go to higher values of temperature and found nothing new. The
plot for the mean energies of both the spin 1 and spin $\frac{1}{2}$
systems is shown in figure 2. The value of cut-off is $L=\frac{59}{2}$.  

\begin{figure}
\begin{center}
\mbox{
    \leavevmode
   { 
      \includegraphics[height=4.5in,width=2.5in, angle=270]{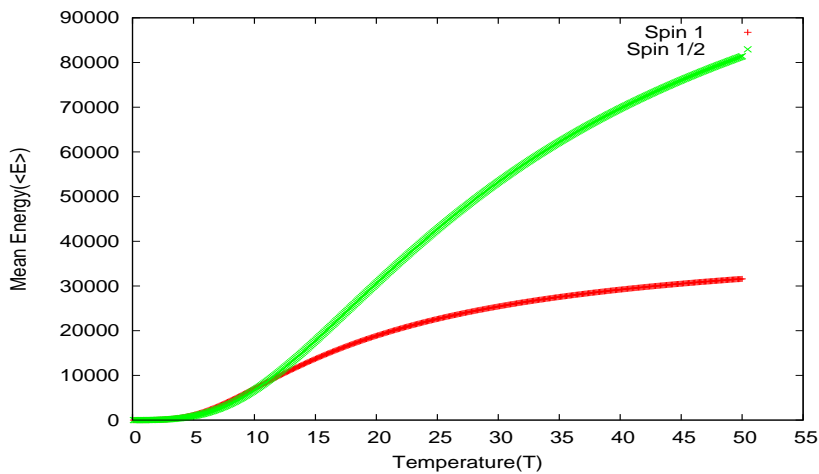} }}
 \end{center}
   \caption{The mean energies of the spin 1 and spin $\frac{1}{2}$ systems.}
\end{figure}

In figure 2, the green curve shows the mean energy for the spin
$\frac{1}{2}$ system as a function of temperature and the red one
shows the corresponding curve for the spin 1 system. The curves
become flat for higher values of temperature. This is due to the
presence of the cut-off angular momentum in our sum. If we go to
higher values of temperature this flattening occurs towards the
higher temperatures considered. The plot clearly shows that the mean
energy of the spin $\frac{1}{2}$ system is much higher than the spin
1 system. This is inspite of the spin 1 system having more number of
degrees of freedom than the spin $\frac{1}{2}$ system. We know of no
such analogous behavior in higher dimensions.

Another interesting feature in the behavior of these curves is the
crossing of the two curves for low values of temperature. This is
not clear in figure 2 but is shown explicitly in figure 3. This plot
shows that the mean energy of the spin $\frac{1}{2}$ system is
smaller than the spin 1 system till about $T=10.77$ after which it
stays above the spin 1 curve.


  \begin{figure}
\begin{center}
\mbox{
    \leavevmode
    { 
      \includegraphics[height=4.5in,width=2.5in, angle=270]{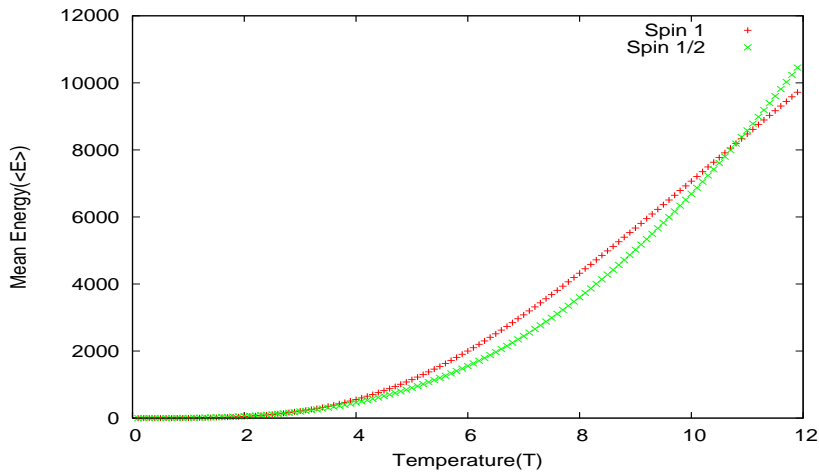} }}
 \end{center}
     \caption{The crossing of the mean energy curves of the spin 1 and spin $\frac{1}{2}$ systems.}
\end{figure}

We now try to explain the cause of this unusual behavior by looking
closely at the distributions of the eigenvalues of the Dirac operators of the two systems.

\subsection{Reasons for the strange behavior}
The main reason can be understood once we look at the spectrum of
the Dirac operator in the two cases. 

Using the expressions for the energy as a function of the degeneracy we can 
study the differences between the two systems. The relation is given by
Eq.(\ref{evghalf}) for the spin $\frac{1}{2}$ case and Eq.(\ref{evgone}) for the
spin 1 case. 

This plot of the energy eigenvalues as a function of their degeneracies is
shown in figure 4.


  \begin{figure}
\begin{center}
\mbox{
    \leavevmode
    { 
      \includegraphics[height=4.5in,width=2.5in, angle=270]{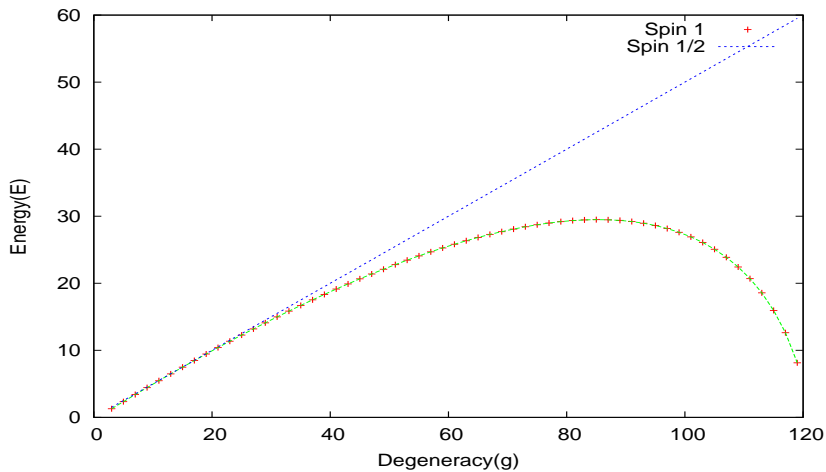} }}
\end{center}
\caption{Energy as a function of the degeneracy $g$ for the spin $\frac{1}{2}$ and spin 1 systems.}
\end{figure}

For a given cut-off $L$, figure 4 clearly indicates that eigenvalues of the spin $\frac{1}{2}$ Dirac operator exceeds that 
of the spin 1 Dirac operator except for small values of the degeneracy $g$. The plot in figure 4
is shown only for positive values of the energy eigenvalue $E$. In the spin $\frac{1}{2}$ case the energy
eigenvalues linearly increase with the degeneracy $g$ and so the maximum eigenvalue occurs
for the $j$ value $2L-\frac{1}{2}$. We have ignored the maximum $j$ value of $2L+\frac{1}{2}$ as they correspond to
unpaired eigenstates of the Dirac operator, which will be inconsistent given the chirality of the spin $\frac{1}{2}$ system
squares to 1. This is a feature of the operator $\vec{\sigma}.\vec{\mathcal{L}}$ in the spin $\frac{1}{2}$ Dirac operator
which is \beq{D_{\frac{1}{2}}= \vec{\sigma}.\vec{\mathcal{L}} +
\frac{1}{2},}\eeq where $\vec{\sigma}$ are the Pauli matrices.

In the spin 1 case the maximum eigenvalue is $L$ and this occurs for some intermediate
value of the degeneracy $g$ as can be seen in figure 4.
The reason why the spin 1 Dirac operator consists of all eigenvalues
ranging from $-L$ to $+L$ for a given $L$ can be seen by looking at
the operator in the continuum given by Eq.(\ref{CD3}) which we write
here again: \beq{D_3 = \vec{\Sigma}.\vec{\mathcal{L}} -
(\vec{\Sigma}.\hat{x})^2 +2}\eeq where $\vec{\Sigma}$
are the matrices of the spin 1 representation of $SU(2)$. The term
$\vec{\Sigma}.\hat{x}$ makes the analytic computation of the
spectrum in the spin 1 case difficult  when compared to the spin
$\frac{1}{2}$ case. We believe this term to be also the cause of the
varied spectrum of the spin 1 system.

\subsection{Zero modes of the spin 1 Dirac operator}

The spectrum of the spin 1 Dirac operator consists of a number of
zero modes for each cut-off angular momentum $L$. The number of such
zero eigenvalues follows a simple power law as a function of $L$. This number was found to
be $(2L+1)^2+2$. This is an exact result and can be found analytically
as explained in~\cite{SdPp}. This has also been verified numerically. With this result it follows immediately that
the number of positive eigenvalues of the spin 1 system are $(2L+1)^2-1$. The spin $\frac{1}{2}$ Dirac operator
has no zero modes as it has non-zero trace. Removing
the states corresponding to the top mode gives us $4L^2+2L$ states with positive eigenvalues. The zero modes in the spin 1 case
drastically reduce the total number of states corresponding to positive eigenvalues to $(2L+1)^2-1$ but this is 
still more than the corresponding number of states  in the spin $\frac{1}{2}$ case. 

The counting of the zero modes and the behavior of the spectrum with degeneracy in the two
cases justify the counter-intuitive behavior of the mean energies. 

We now digress a bit to remark about the plot in figure 4. We try to speculate the energy versus degeneracies curves for
higher spin Dirac operators. To do this we first find the number of zero modes for higher spin Dirac operators.  We will compute this for 
the integer spin case.   

To construct higher spin Dirac operators on $S_F^2$, we need to construct operators acting on $Mat(2L+1)\otimes \mathbb{C}^{2k+1}$
where $k$ is the desired spin. The spectrum of these operators will in general be hard to compute due to the presence of $\vec{\Sigma}.\hat{x}$
terms just as in the spin 1 case.

The analytic computation of the number of zero modes was given in~\cite{SdPp}. We extend those
arguments to higher spins in the following. Consider the spectrum of the total angular momentum $j$ for a given cut-off $L$: \beq{\textrm{Spec}~ \vec{J} \in \{0,1, 2,\cdots , 2L-k, \cdots , 2L, \cdots , 2L+k \}.}\eeq For an even-integer spin $k$, the number of zero-modes can be found 
by computing the following sum \beq{\sum_{j=0}^{2L-k}(2j+1) + \sum_{j=-\frac{k-2}{2}}^{\frac{k}{2}}\left[2(2L+2j)+1\right] = (2L+1)^2+k^2+k.}\eeq
For an odd-integer spin $k$, this number is \beq{\sum_{j=0}^{2L-k-1}(2j+1) + \sum_{j=-\frac{k+1}{2}}^{\frac{k-1}{2}}\left[2(2L+2j+1)+1\right] = (2L+1)^2+k^2+k.}\eeq These computations hold as there exists a traceless Dirac operator for all integer spin Dirac operators on $S_F^2$. This
is because we can construct an integer spin Dirac operator from the following combination of generators of GW algebra~\cite{ApbPp}: \beq{D_k = L\left(\frac{\Gamma_L^L - \Gamma_L^R}{2}\right).}\eeq As $tr(\Gamma_L^L)=tr(\Gamma_L^R)$, this operator is traceless.

In the case of the Dirac operators for half-integral spins, there exists no such combinations of generators of GW algebras which have 0 trace.
This makes the number of states with positive energy eigenvalues for spin $k$ and spin $k-\frac{1}{2}$, for integer $k$, comparable. 

We can then go on to compute their mean energies and compare them. We suspect $\langle E_{k-\frac{1}{2}}\rangle >\langle E_k\rangle$ to hold but we have no analytic proof for this. We could however compute the spectrums for the two Dirac operators numerically and carry out this comparison, but we do not do this here and save it for future work. 

The reason why this is interesting is the following. It seems from the plot in figure 4 that the
behavior of $\frac{E}{g}$ for small values of $g$ is similar for higher spins as well. We leave this as a conjecture as we have no analytic proof for this but do have strong reasons to suspect so. 

It is also very likely that the plots of $E$ versus $g$ for higher spins will fall below the $E=\frac{g}{2}$ curve. This is expected due to the fact that higher spin Dirac operators contain $\vec{\Sigma}.\hat{x}$ terms along with $\vec{\Sigma}.\vec{\mathcal{L}}$~\cite{ApbPp}. $\vec{\Sigma}$ is the $2k+1$ dimensional representation of $SU(2)$ for some spin $k$. The $\vec{\Sigma}.\hat{x}$ term disrupts the linearity between the energy and degeneracy. It is easy to see this as a linear relation between the energy and the degeneracy is only possible for a Dirac operator which has just a $\vec{\Sigma}.\vec{\mathcal{L}}$ term apart from constant terms. This can be seen analytically for any given spin $k$ by looking at the spectrum of $\vec{\Sigma}.\vec{\mathcal{L}}$:\beq{\textrm{Spec} ~\vec{\Sigma}.\vec{\mathcal{L}} = (k-m)(2j-k+m) - k^2 -m ~~ m\in\{0,1,\cdots , 2k\}.}\eeq Only the spin $\frac{1}{2}$ Dirac operator contains just the $\vec{\sigma}.\vec{\mathcal{L}}$ term leading to the linear relation between its energy and their multiplicities. 

The $\vec{\Sigma}.\hat{x}$ terms are present in the spin 1 case and it was remarked that these terms cause the energy versus degeneracy curve in figure 4. As these terms also occur for higher spin Dirac operators we expect a similar behavior from these systems. The reason why they occur for all higher spin Dirac operators is because of the fact that the fuzzy versions of these higher spin Dirac operators contain terms of the form $$\frac{(\vec{\Sigma}.\vec{L^L})^n - (\vec{\Sigma}.\vec{L^R})^n}{L^{n-1}}.$$ The continuum limit of these terms contain $\vec{\Sigma}.\hat{x}$ terms. This is explained in detail in~\cite{ApbPp}.  

The preceding statements prove the non-linearity between the energy and their degeneracies for all Dirac operators other than the spin $\frac{1}{2}$ system. They however do not show that these curves fall below the corresponding curve for the spin $\frac{1}{2}$ system. A complete answer to this question would only come from a numerical analysis of this system and at present we leave this question as a worthy one to explore in the future.


\subsection{Specific heats of the two systems}

The specific heat is defined as the derivative of the mean energy
with respect to temperature. A straightforward computation gives the
specific heat as \beq{C_v = \frac{1}{T^2}\sum_i \frac{g_iE_i^2e^{\beta
E_i}}{(e^{\beta E_i}+1)^2}.}\eeq As expected here too we find the
specific heat of the spin $\frac{1}{2}$ system to be more than that
of the spin 1 system. This is shown in figure 5.  

There is a region till $T=10.77$ where the specific heat of the spin 1 system is more
than that of the spin $\frac{1}{2}$ system. This can be seen as a result of the crossing
of the mean energy curves for the two systems as shown in figure 3.

 \begin{figure}
\begin{center}
\mbox{
    \leavevmode
    { 
      \includegraphics[height=4.5in,width=2.5in, angle=270]{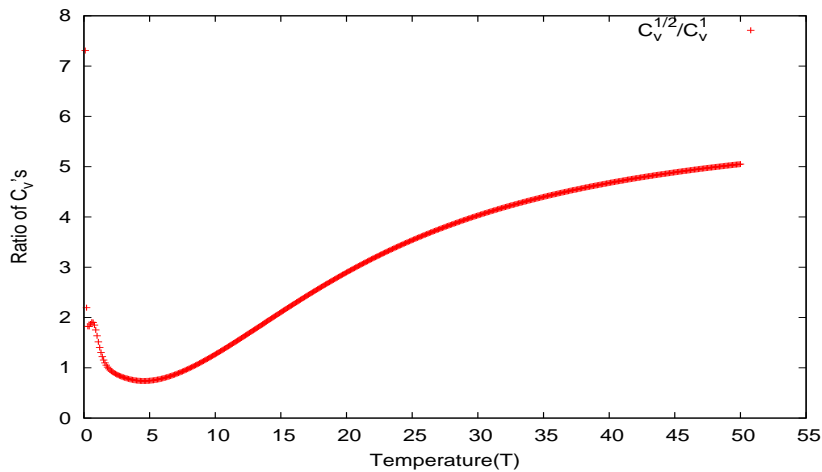} }}
\end{center}
\caption{Ratio of specific heats $\frac{C_v^{\frac{1}{2}}}{C_v^1}$}
\end{figure}

\subsection{Entropies of the two systems}
 The entropy is given by the equation \beq{S = \sum_i g_iln\left(1+e^{-\beta E_i}\right) + \frac{1}{T}\sum\frac{E_i}{1+e^{-\beta E_i}}.}\eeq 
This follows from \beq{S = ln Z + \beta \langle E\rangle .}\eeq From these formulas it can be easily seen that the entropy of a spin 1 system is less than that of a spin $\frac{1}{2}$ system. 

\section{Numerical Results: II}
\subsection{Deviations from the Ideal Gas Law}

For a system of non-interacting massless particles obeying Fermionic
statistics, the mean energy goes as $T^4$ in 3+1 dimensions. This can be seen as 
follows: \beq{\langle E\rangle = \int d^3p\frac{p}{e^{\beta p}+1}}\eeq where $p$ is the 
energy of the massless particle. This is the dispersion law
for a massless particle on a flat space which has the Poincare group has its
group of symmetries. We now substitute $$x=\frac{p}{T}$$ to find 
\beq{\langle E\rangle \propto T^4\int_0^{\infty} dx\frac{x^3}{e^x+1}.}\eeq    
In a similar manner, in 2+1 dimensions, the mean energy goes as
$T^3$. This law holds however only for a system living on a flat 2+1 dimensional
spacetime. 

In the case of the spin $\frac{1}{2}$ system living on $S^2$ or $S_F^2$ the mean energy
is given by \beq{\langle E_{\frac{1}{2}}\rangle = \sum_{j=0}^{2L+\frac{1}{2}} \frac{1}{2} \frac{(2j+1)^2}{e^{\frac{2j+1}{2T}}+1}.}\eeq
We have cut-off the sum with a cut-off $L$. If we arbitrarily increase the value of the cut-off $L$ we will find the sum replaced
by an integral over $j$ and the upper limit in the sum goes to $\infty$. In the above equation make the substitution 
\beq{\frac{2j+1}{T} = x.}\eeq This makes the sum \beq{\langle E_{\frac{1}{2}}\rangle = \frac{T^2}{2} \sum_{x=\frac{1}{T}}^{\frac{4L+2}{T}} 
\frac{x^2}{e^{\frac{x}{2}}+1}.}\eeq As the limits of the sum depend on the temperature $T$ we get no definite relation between 
the mean energy and temperature. It should be noted that the upper limit is dependent on $T$ due to the cut-off $L$. We can remove
this by allowing $L$ to go to $\infty$. In such a case, as already mentioned the sum becomes an integral making the dependence go
as $T^3$. This still does not remove the $T$ dependence from the lower limit of the integral. This is due to the dispersion relation
for the spin $\frac{1}{2}$ particle which goes as $j+\frac{1}{2}$. The additional $\frac{1}{2}$ can be attributed to the curvature of the sphere
the system lives on. This results in the deviation from ideal gas law on 2+1 dimensional space.

Similar arguments hold for the spin 1 case also. This can be easily seen from our analytic expressions for the spectrum of the spin 1 Dirac operator as seen in the previous section. We do not write the simple details of this here. 

The deviations for the spin $\frac{1}{2}$ and the spin 1 system are shown in the plots in figures 6 and 7.


 \begin{figure}
\begin{center}
\mbox{
    \leavevmode
    { 
      \includegraphics[height=2.5in,width=4.5in, angle=0]{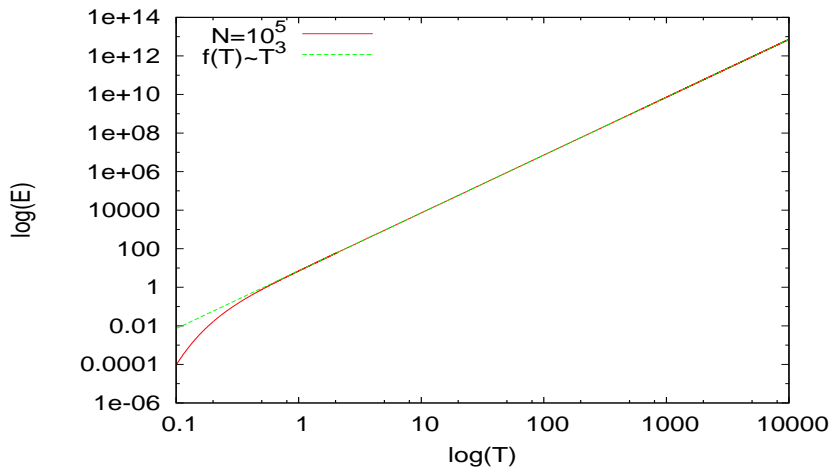} }}
 \end{center}
     \caption{Deviation from ideal gas law for the spin $\frac{1}{2}$ system.}
\end{figure}


 \begin{figure}
\begin{center}
\mbox{
    \leavevmode
    { 
      \includegraphics[height=2.5in,width=4.5in, angle=0]{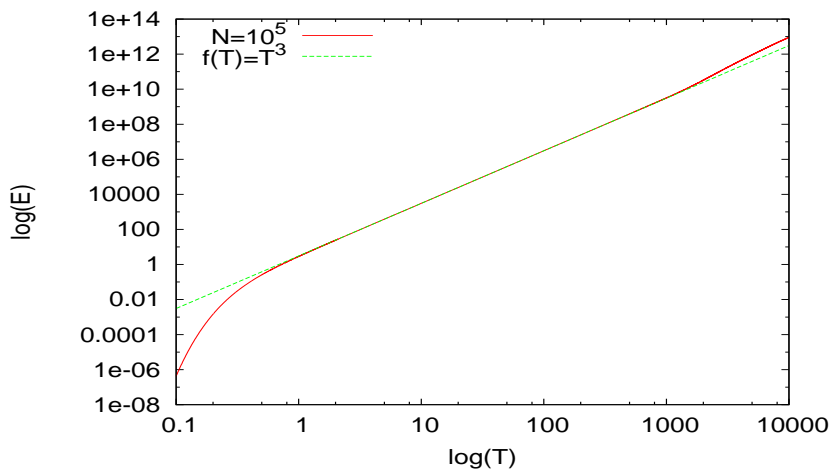} }}
\end{center}
\caption{Deviation from the ideal gas law for the spin 1 system.}
\end{figure}


\section{Conclusions}
Spin systems on the fuzzy 2-sphere show several interesting features
as discussed in this paper. The main result being the mean energy of
a spin $\frac{1}{2}$ system is greater than that of a spin 1 system.
This is a novel phenomenon with no known higher dimensional analog.
Its implications are worth exploring. The simple power law of the
zero modes is another remarkable feature of these systems. 

Speculations were made about the spectrums of higher spin Dirac operators
on $S_F^2$. Though we could not arrive at a concrete result, we conjectured
that it is likely that the mean energy of the spin $\frac{1}{2}$ system will be
more than that of the mean energies of other spins greater than $\frac{1}{2}$.
This gives an upper bound on the mean energy of these spin systems
on $S_F^2$. This bound will also translate to the entropies of these
systems, thus making the entropies of these spin systems to be bounded by 
the entropy of the spin $\frac{1}{2}$ system. This would mean that an increase
in the number of states on $S_F^2$ have no effect on the entropy, a property
very much reminiscent of the holographic principle though there is no bulk theory 
involved here. This is conjectural and we plan to explore this in the future.

The deviation from the ideal gas law is another interesting
phenomenon in this fuzzy system. It can definitely serve as a model
for explaining 2 dimensional systems which show such behavior.

We carried out similar computations by assuming other exotic
statistics like anyonic statistics, but found nothing interesting.
For this we used the following formula instead of Eq.(\ref{spin1en})
\beq{\langle E\rangle = \sum_i \frac{E_i}{e^{\beta E_i}+\alpha}}\eeq
where $\alpha$ is allowed to vary from $-1\dots, 0, \dots +1$ going
through all possible statistics between bosons and fermions. However
there are different approaches in formulating the statistical
mechanics of anyons and our treatment is by no means complete. We do
plan to study this further in the future as we maybe missing some
connection. (See also~\cite{Jd} in this regard.)

Dirac operators appear in 2-dimensional graphene systems in
condensed matter physics. It would be interesting to know if the
continuum limits of our models will be useful for similar carbon
systems which have a spherical shape like fullerene ($C_{60}$). We
do not know of any direct application now.

Dirac operators can be constructed on other fuzzy spaces as well. It
is a very interesting problem to see if the method of GW algebras
can be used to construct higher spin Dirac operators on other fuzzy
spaces. A study of their thermodynamics could lead to a lot of rich
and strange phenomena like the ones found in this paper.

These can have an application to cosmology in terms of the thermal history of the universe.  We will report
these results in a forthcoming paper~\cite{Future}. Also see
in this regard the paper~\cite{WangWang} where an interesting toy model of the universe
as a fuzzy sphere was presented.

The new results found here are encouraging, and a hunt for
applications of these features of fuzzy systems is a worthy
endeavor.

\section{Acknowledgements}
 We thank Prof.A.P.Balachandran, Prof.T.R.Govindrajan and Prof. R. Shankar
for useful discussions and references. We are grateful to Prof.Marco
Panero for useful comments on our earlier work~\cite{SdPp}. PP
thanks Vinu Lukose and A.B.Belliappa for helping in coding and
preparing the manuscript. 
 PP is grateful to Prof.T.R.Govindarajan for the wonderful hospitality at IMSc,
Chennai. PP was supported by DOE under the grant number
DE-FG02-85ER40231.

\end{document}